\documentclass{jpconf}

\usepackage{pstricks,amssymb,bm}

\def\diag{{\mathop{\mathrm{diag}}}}
\newtheorem{theorem}{Theorem}

\begin{document}
\title{Re`class'ification of `quant'ified classical simulated annealing}
\author{Toshiyuki Tanaka}
\address{Department of Systems Science, Graduate School of Informatics, Kyoto University,\\
36-1 Yoshida Hon-machi, Sakyo-ku, Kyoto-shi, Kyoto 606-8501, Japan}
\ead{tt@i.kyoto-u.ac.jp}

\begin{abstract}
We discuss a classical reinterpretation of 
quantum-mechanics-based analysis of classical Markov chains 
with detailed balance, that is based on the quantum-classical correspondence.  
The classical reinterpretation is then used to demonstrate 
that it successfully reproduces a sufficient condition 
for cooling schedule in classical simulated annealing, 
which has the inverse-logarithm scaling.  
\end{abstract}

\section{Introduction}
Theoretical properties of simulated annealing~\cite{KirkpatrickGellatVecchi1983} 
have been extensively studied in the 1980s~\cite{vanLaarhovenAarts1988,AartsKorst1989}.  
One of the main issues in those research activities was 
regarding the annealing schedule: 
How should one decrease temperature $T(t)$ as a function of time $t$ 
in order to finally arrive at a globally optimum solution?  
Geman and Geman~\cite{GemanGeman1984} were the first to obtain 
an answer, which states a sufficient condition of the form $T(t)\ge O(1/\log t)$.  
The inverse-logarithm scaling turned out to be universal, 
in the sense that it is also sufficient for many variants of 
simulated annealing and some other algorithms.  
Hajek~\cite{Hajek1988} proved a necessary and sufficient condition 
which also has the inverse-logarithm form, 
showing that one cannot do the cooling any faster than that 
while guaranteeing global optimality. 

Somma et al., in their recent contribution~\cite{SommaBatistaOrtiz2007}, 
have shown that the inverse-logarithm scaling of simulated annealing 
can also be obtained via the adiabatic condition~\cite{Messiah1960} 
of a related quantum-mechanical system.  
The relationship between the original Markov chain in simulated annealing 
and the quantum system is established via the so-called 
classical-quantum mapping or quantum-classical correspondence%
~\cite{Henley2004,CastelnovoChamonMudryPujol2005,VerstraeteWolfPerezGarciaCirac2006}.  
In this paper, we discuss a classical reformulation of quantum equivalent 
of a classical Markov chain with detailed balance, in order to elucidate 
mathematical structure of the correspondence between a Markov chain and 
its quantum equivalent, without making reference to quantum mechanics.  
We also discuss another classical reformulation of the argument 
deriving the optimal inverse-logarithm scaling 
of annealing schedule~\cite{SommaBatistaOrtiz2007}  
(see also~\cite{MoritaNishimori2008pre}), 
which is based on the quantum adiabatic theorem.  

This paper is organized as follows.  
In section 2 we first provide a basic formulation of classical Markov chains 
with detailed balance, and derive its $\alpha$-representation.  
A local linear approximation of the time evolution in terms of $\alpha$-representation 
is also discussed.  
Our derivation of the inverse-logarithm scaling of simulated annealing 
is discussed in section 3.  
A ``chasing'' view of simulated annealing, 
that is based on the local linear approximation based on the 0-representation, 
and a bound of the largest negative eigenvalue 
are used in the derivation.  
In section 4 we discuss relation between our formulation 
and the stochastic matrix form decomposition, 
which is defined and discussed extensively in~\cite{CastelnovoChamonMudryPujol2005}.  
Section 5 concludes the paper.  

\section{Basic formulations}
\subsection{Markov chains}

Let $S$ denote a state space, which is a finite set of cardinarity $N$. 
Let $E$ be an ``energy function'' defined on $S$, 
which associates a state $i\in S$ with its energy $E_i$. 
Then, one can define a probability distribution on $S$, 
in terms of a probability vector $\bar{\bm{\rho}}=(\bar{\rho}_i)$, as 
\begin{equation}
\label{eq:GBdist}
\bar{\rho}_i=\frac{e^{-\beta E_i}}{Z},\quad Z=\sum_{i\in S}e^{-\beta E_i},
\end{equation}
which is the Gibbs-Boltzmann distribution induced by the 
energy function $E$, with $\beta>0$ a parameter corresponding to the inverse temperature. 

Let us consider a undirected graph $G$ with $S$ its vertex set 
and an edge set $L$.  We assume $G$ to be a connected graph, 
without self-edge (loop).  
We define a transition matrix $M=(m_{ij})$ as 
\begin{equation}
m_{ij}=\left\{
\begin{array}{cc}
w_{ij}e^{\beta(E_j-E_i)/2} & ((ij)\in L)\\
-\sum_{k:\,(ik)\in L}m_{ki} & (i=j)\\
0 & (\hbox{otherwise})
\end{array}\right.,
\end{equation}
where $W=(w_{ij})$ is a symmetric matrix with 
$w_{ij}>0$ for $(ij)\in L$.  
On the basis of the transition matrix $M$, one can define a 
continuous-time Markov chain, as 
\begin{equation}
\label{eq:MC}
\dot{\bm{\rho}}=M\bm{\rho}.
\end{equation}
The connectedness of the graph $G$ induces irreducibility 
of the Markov chain.  
The Markov chain is also aperiodic, so that it is ergodic, 
and therefore bears a unique equilibrium distribution. 
The Gibbs-Boltzmann distribution~(\ref{eq:GBdist}) is the equilibrium distribution 
of the Markov chain, since $M\bar{\bm{\rho}}={\bf 0}$ holds. 

The formulation presented here is general, 
including various typical systems as special cases.  
For example, conventional Ising spin systems are described 
by letting $S=\{-1,\,1\}^n$ with $N=2^n$ 
and $L$ having an $n$-dimensional hypercubic structure.  
Metropolis and Glauber dynamics are implemented by letting 
\begin{equation}
w_{ij}\propto\max\left\{e^{\beta(E_j-E_i)/2},\ e^{\beta(E_i-E_j)/2}\right\},
\end{equation}
and 
\begin{equation}
w_{ij}\propto\frac{1}{e^{\beta(E_j-E_i)/2}+e^{\beta(E_i-E_j)/2}},
\end{equation}
respectively, as mentioned in~\cite{CastelnovoChamonMudryPujol2005}.

\subsection{$\alpha$-representation}
We discuss a different representation of the continuous-time Markov chain~(\ref{eq:MC}), 
in view of the classical-to-quantum mapping utilized in~\cite{SommaBatistaOrtiz2007}.  
Although the quantum reformulation mapped from a classical Markov chain makes use of 
square roots of probabilities $\{\sqrt{\rho_i}\}$, 
we here discuss a slightly more generalized expression 
which is based on the so-called $\alpha$-representation 
of $\bm{\rho}$.  

\noindent
{\bf Definition}. We define the $\alpha$-representation 
$\bm{\psi}^{(\alpha)}=(\psi^{(\alpha)}_i)$ of $\bm{\rho}$ as 
\begin{equation}
\label{eq:alpha-rep}
\psi^{(\alpha)}_i=\frac{2}{1-\alpha}\rho_i^{(1-\alpha)/2}.
\end{equation}

The concept of $\alpha$-representation is originally 
introduced in information geometry~\cite{Amari1985,AmariNagaoka2000}, 
in order to discuss intrinsic geometrical structures 
of statistical manifolds. 
Taking square roots of probabilities 
corresponds to considering 0-representation.  
Although not used in this paper, 1-representation is defined as 
\begin{equation}
\psi_i^{(1)}=\log\rho_i.
\end{equation}

We next derive an expression of the Markov chain 
in terms of the $\alpha$-representation.  
One has 
\begin{eqnarray}
\dot{\psi}_i^{(\alpha)}
&=&\rho_i^{-(1+\alpha)/2}\sum_{j\in S}m_{ij}\rho_j
\nonumber\\
&=&\rho_i^{-(1+\alpha)/2}\frac{1-\alpha}{2}\sum_{j\in S}
m_{ij}\rho_j^{(1+\alpha)/2}\psi^{(\alpha)}_j,
\end{eqnarray}
which is rewritten, in a vector-matrix form, as 
\begin{equation}
\dot{\bm{\psi}}^{(\alpha)}
=\frac{1-\alpha}{2}H^{(-\alpha)}\bm{\psi}^{(\alpha)},
\end{equation}
where the matrix $H^{(\alpha)}$ is defined as 
\begin{equation}
\label{eq:mtxH}
H^{(\alpha)}=(\Psi^{(\alpha)})^{-1}M\Psi^{(\alpha)},
\end{equation}
with $\Psi^{(\alpha)}=\diag(\psi_i^{(\alpha)})$.  
Clearly, eigenvalues of the matrix $H^{(\alpha)}$ 
are the same as those of $M$.  
The elements of the matrix $H^{(-\alpha)}=(h_{ij}^{(-\alpha)})$ are given by 
\begin{equation}
h_{ij}^{(-\alpha)}=\left\{
\begin{array}{cc}
w_{ij}e^{\beta(E_j-E_i)/2}\left(\psi_i^{(-\alpha)}\right)^{-1}\psi_j^{(-\alpha)} & ((ij)\in L)\\
-\sum_{k:\,(ik)\in L}w_{ki}e^{\beta(E_i-E_k)/2} & (i=j)\\
0 & (\hbox{otherwise})
\end{array}\right.,
\end{equation}
and consequently, 
\begin{equation}
\bar{h}_{ij}^{(-\alpha)}=h_{ij}^{(-\alpha)}|_{\bm{\rho}=\bar{\bm{\rho}}}=\left\{
\begin{array}{cc}
w_{ij}e^{\alpha\beta(E_i-E_j)/2} & ((ij)\in L)\\
-\sum_{k:\,(ik)\in L}w_{ki}e^{\beta(E_i-E_k)/2} & (i=j)\\
0 & (\hbox{otherwise})
\end{array}\right..
\end{equation}
The above expression evidently shows 
that the 0-representation is special in our formulation, 
in that the matrix $\bar{H}^{(-\alpha)}=H^{(-\alpha)}|_{\bm{\rho}=\bar{\bm{\rho}}}$ 
becomes symmetric when $\alpha=0$, that is, 
under the 0-representation.  
The fact that the 0-representation symmetrizes the transition matrix $M$ 
was also mentioned in~\cite{MukherjeeNakanishiFuchs1994}, 
in order to state that eigenvalues of $M$ are all real.  
It should be noted that the matrix $H^{(-\alpha)}$ is 
dependent on $\bm{\psi}^{(\alpha)}$ via $\Psi^{(-\alpha)}$ and therefore 
$H^{(0)}$ does not symmetric at $\bm{\rho}\not=\bar{\bm{\rho}}$ in general. 

\subsection{Time evolution}
We discuss linearization of the $\alpha$-representation 
of the dynamical equation.  
Starting from the nonlinear dynamics 
\begin{equation}
\dot{\psi}_i^{(\alpha)}
=\frac{1-\alpha}{2}
\left(\psi_i^{(\alpha)}\right)^{-(1+\alpha)/(1-\alpha)}
\sum_{j\in S}m_{ij}
\left(\psi_j^{(\alpha)}\right)^{2/(1-\alpha)},
\end{equation}
and considering a small perturbation $\delta\bm{\psi}^{(\alpha)}$ 
around $\bm{\psi}^{(\alpha)}$, 
we obtain the following linearized system 
which describes time evolution of $\delta\bm{\psi}^{(\alpha)}$: 
\begin{eqnarray}
\label{eq:linearization}
\delta\dot{\psi}_i^{(\alpha)}
&=&
\left(\psi_i^{(\alpha)}\right)^{-(1+\alpha)/(1-\alpha)}
\sum_{j\in S}m_{ij}
\left(\psi_j^{(\alpha)}\right)^{(1+\alpha)/(1-\alpha)}
\delta\psi^{(\alpha)}_j
\nonumber\\
&&{}-\frac{1+\alpha}{2}\left[
\left(\psi_i^{(\alpha)}\right)^{-2/(1-\alpha)}
\sum_{j\in S}m_{ij}
\left(\psi_j^{(\alpha)}\right)^{2/(1-\alpha)}
\right]\delta\psi^{(\alpha)}_i+o\left(\|\delta\bm{\psi}^{(\alpha)}\|\right).
\end{eqnarray}
In particular, observing that the second term of the right-hand side 
of~(\ref{eq:linearization}) vanishes 
at the Gibbs-Boltzmann distribution $\bar{\bm{\rho}}$, 
irrespective of the value of $\alpha$, 
the linearization around the equilibrium point becomes, 
ignoring higher-order terms, 
\begin{equation}
\label{eq:lin_at_GB}
\delta\dot{\bm{\psi}}^{(\alpha)}
=\bar{H}^{(-\alpha)}\delta\bm{\psi}^{(\alpha)}.
\end{equation}
Equation~(\ref{eq:lin_at_GB}) states that 
the matrix $\bar{H}^{(-\alpha)}$ governs the local dynamics 
described in terms of $\alpha$-representation 
in the vicinity of the equilibrium distribution $\bar{\bm{\rho}}$.  
It should be noted that the right-hand side of~(\ref{eq:linearization}) 
is in general not a projection of $H^{(-\alpha)}\delta\bm{\psi}^{(\alpha)}$ onto 
the manifold of probability distributions in $\alpha$-representation, 
defined as 
\begin{equation}
\sum_{i\in S}\left(\frac{1-\alpha}{2}\psi_i^{(\alpha)}\right)^{2/(1-\alpha)}=1.
\end{equation}

\section{Simulated annealing}
\subsection{Relaxation in annealing}
With the inverse temperature $\beta$ fixed, 
the distribution following the Markov chain 
relaxes toward the Gibbs-Boltzmann distribution.  
The basic idea behind simulated annealing is that 
by gradually reducing the temperature one can arrive 
at a distribution which concentrates on a set 
of minimum-energy states.  
Thus, by performing simulations of the Markov chain 
with a proper cooling schedule, one expects to obtain 
minimum-energy states with probability close to 1.  
One of the basic questions regarding simulated annealing 
is to determine the cooling schedule which guarantees 
convergence to minimum-energy states.  

We wish to study this problem via the linearized local dynamics 
in $\alpha$-representation~(\ref{eq:lin_at_GB}), with $\alpha=0$.  
Intuitively, our expectation is that 
if simulated annealing works well 
the distribution should stay very close to 
instantaneous equilibrium distributions 
as $\beta$ is changed slowly enough.  
If it is the case, then arguments that are based on 
the local linear approximation around the equilibrium~(\ref{eq:lin_at_GB}) 
will be justified.  
Since the coefficient matrix $\bar{H}^{(0)}$ is symmetric, 
all eigenvalues are real, so that 
the local dynamics around equilibrium is a simple 
linear relaxation toward the equilibrium, 
with negative eigenvalues of $\bar{H}^{(0)}$ govern 
the speed of relaxation.  
In simulated annealing the instantaneous equilibium distribution is 
also slowly drifting as $\beta$ changes.  
One can therefore expect to obtain a minimum-energy distribution 
only if the drift is slow enough so that 
the relaxation process is managed to catch up with the drift. 
What is important for successful convergence of simulated annealing 
is thus the largest negative eigenvalue of $\bar{H}^{(0)}$.  

\subsection{Bound on largest negative eigenvalue}
We let 
\begin{equation}
{\cal M}=(bI+\chi\bar{H}^{(0)})^N,
\end{equation}
where $\chi=e^{-\beta d/2}/w_{\rm max}$, with $d=\max_{i,\,j}|E_i-E_j|$ 
and $w_{\rm max}=\max_{(ij)\in L}w_{ij}$, 
is to make diagonal elements of $\chi\bar{H}^{(0)}$ nondiverging as $\beta$ gets large, 
and where 
\begin{equation}
b=1+\max_{i\in S}\sum_{k:(ki)\in L}\frac{w_{ki}}{w_{\rm max}}
\end{equation}
is chosen so that $bI+\chi\bar{H}^{(0)}$ becomes a non-negative matrix.  
Irreducibility of the original Markov chain guarantees 
${\cal M}$ to be a (strictly) positive matrix.  

The following theorem for positive matrices, 
due to Hopf~\cite{Hopf1963a} in its operator form, 
is applied to obtain an upper bound 
of the largest negative eigenvalue.  

\begin{theorem}
Let $A=(a_{ij})$ be a square matrix that is positive, i.e., $a_{ij}>0$ holds for all $i,\,j$.  
Then the maximum eigenvalue $\lambda_0$ of $A$ and any other eigenvalues $\lambda$ 
satisfy the inequality 
\begin{equation}
|\lambda|\le\frac{\kappa-1}{\kappa+1}\lambda_0,
\end{equation}
where 
\begin{equation}
\kappa=\max_{i,\,j,\,k}\frac{a_{ik}}{a_{jk}}.
\end{equation}
\end{theorem}

All positive elements of the matrix $bI+\chi\bar{H}^{(0)}$ are 
bounded from below by $\min\{1,\,w_{\rm min}\chi\}$, 
where $w_{\rm min}=\min_{(ij)\in L}w_{ij}$, 
and $w_{\rm min}\chi$ actually gives the lower bound 
for not too small values of $\beta$.  
A lower bound of the minimum element of ${\cal M}$ 
is thus $(w_{\rm min}\chi)^N$.  
Alternatively, the matrix $bI+\chi\bar{H}^{(0)}$ is 
upper bounded componentwise by the matrix 
$(b-1)I+{\bf 11}^T$, where ${\bf 11}^T$ is an all-1 matrix, 
so that an upper bound of the maximum element of ${\cal M}$ 
is given by $(3N)^N$.  
An upper bound of the parameter $\kappa$ is therefore 
evaluated as 
\begin{equation}
\kappa\le\left(\frac{3N}{w_{\rm min}\chi}\right)^N.
\end{equation}
Note that symmetry of the matrix $bI+\chi\bar{H}^{(0)}$, 
and hence of ${\cal M}$, makes the argument of bounding $\kappa$ 
straightforward, thereby demonstrating efficiency 
of the 0-representation.  

Let $\lambda$ be a negative eigenvalue of $\bar{H}^{(0)}$. 
Since we know that $\bar{H}^{(0)}$ has a zero eigenvalue 
which is the largest, applying theorem 1 yields 
\begin{equation}
(b+\chi\lambda)^N\le\frac{\kappa-1}{\kappa+1}b^N,
\end{equation}
and consequently, 
\begin{equation}
\lambda\le-\frac{2b}{N(\kappa+1)}
\le-\frac{2b(w_{\rm min}\chi)^N}{N[(3N)^N+(w_{\rm min}\chi)^N]}
\le-\frac{b(w_{\rm min}\chi)^N}{N(3N)^N},
\end{equation}
where we used the inequality 
$1-[(\kappa-1)/(\kappa+1)]^{1/N}\ge2/[N(\kappa+1)]$ for $\kappa,\,N\ge1$.  
To make clear its dependence on $\beta$, 
we rewrite it as 
\begin{equation}
\lambda\le -\delta e^{-\beta N(d+d')/2}
,\quad
\delta=\frac{b}{N(3N)^N},
\end{equation}
where we have taken into account possible dependence of $w_{ij}$ on $\beta$, 
by assuming that 
\begin{equation}
\frac{w_{\rm min}}{w_{\rm max}}\ge e^{-\beta d'/2}
,\quad d'\ge 0
\end{equation}
holds.  

\subsection{Simulated annealing as a chase of target}
From now on we assume the inverse temperature $\beta$ to be 
a function of time $t$, and consider speed of drift 
of the instantaneous equilibrium distribution $\bar{\bm{\psi}}^{(0)}$.  
We have 
\begin{equation}
\Bigl\|\dot{\bar{\bm{\psi}}}^{(0)}\Bigr\|^2
=\mathop{\mathrm{Cov}}(E)\dot{\beta}^2
\le C^2e^{-\beta g}\dot{\beta}^2,
\end{equation}
where $g$ is an energy gap between the lowest and the second lowest 
energies in $\{E_i;\,i\in S\}$, and where $C>0$ is a constant 
independent of $\beta$.  

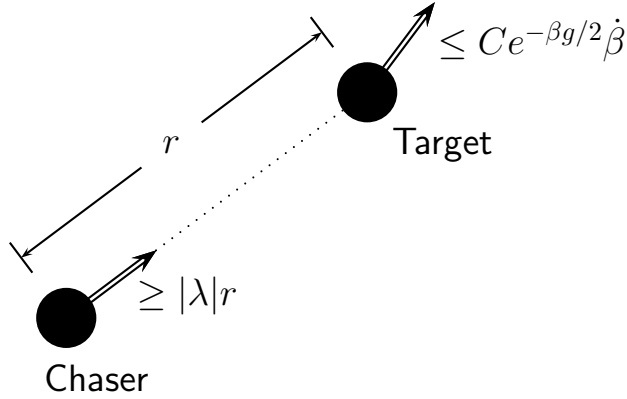
\begin{figure}
\psset{unit=0.1mm}
\begin{center}
\begin{pspicture}(0,0)(800,600)
\psline[linestyle=dotted]{-}(100,150)(500,450)
\rput(-60,80){%
  \psline{-}(85,170)(115,130)
  \psline{<-}(100,150)(260,270)
  \rput(300,300){\large $r$}
  \psline{->}(340,330)(500,450)
  \psline{-}(485,470)(515,430)}
\rput(100,150){%
  \psline[doubleline=true]{->}(0,0)(120,90)%
  \pscircle[fillstyle=solid,fillcolor=black]{40}%
  \rput[l](80,30){\large${}\ge|\lambda|r$}%
  \rput(40,-80){\large\sf Chaser}
}
\rput(500,450){%
  \psline[doubleline=true]{->}(0,0)(90,120)%
  \pscircle[fillstyle=solid,fillcolor=black]{40}%
  \rput[l](80,70){\large${}\le Ce^{-\beta g/2}\dot{\beta}$}
  \rput(100,-70){\large\sf Target}
}
\end{pspicture}
\end{center}
\caption{A ``chasing'' view of simulated annealing.}
\label{fig:chase}
\end{figure}

Now the problem is recast into the problem 
of ``chasing'' a drifting target (see figure~\ref{fig:chase}), 
whose velocity is no more than $Ce^{-\beta g/2}\dot{\beta}$.  
The speed of the chaser is no less than $|\lambda|r$, 
where $r$ is the ``distance'' between the chaser and the target, 
because the speed is determined by gradient-descent 
of a potential surface induced by $\bar{H}^{(0)}$.  
In view of the adiabatic theorem, 
which lays the basis of the quantum-mechanics-based analysis 
of simulated annealing~\cite{SommaBatistaOrtiz2007}, 
we assume that $r$ is small throughout the process, 
so that the local linear approximation of the dynamics is valid.  
We wish to obtain a sufficient condition for $\beta$, 
as a function of time $t$, such that $r$ tends to 0 as $t\to\infty$ and $\beta\to\infty$.  
With a modest amount of foresight, 
we assume that $r$ approaches zero as $r\sim r_0t^{-\gamma}$ with $0<\gamma<1$.  
Since the speed of the chaser should be larger than that of the target, 
as a sufficient condition one has 
\begin{equation}
\delta r_0e^{-\beta N(d+d')/2}t^{-\gamma}>Ce^{-\beta g/2}\dot{\beta}.
\end{equation}
Solving it for $\beta$, we obtain for large enough $t$, 
\begin{equation}
\beta<\frac{2(1-\gamma)}{N(d+d')-g}\log t+O(1).
\end{equation}
For consistency, the difference of the speeds of the chaser and the target 
is equal to $\dot{r}$, which should scale as $t^{-\gamma-1}$, 
yielding $\gamma=g/[N(d+d')]$.  
Collecting these results and ignoring non-dominant terms, 
one finally obtains 
\begin{equation}
\beta^{-1}=T(t)>\frac{N(d+d')}{2\log t}
\end{equation}
as a sufficient condition for simulated annealing 
to converge to a minimum-energy distribution.  

\section{Stochastic matrix form decomposition}
The stochastic matrix form (SMF) decomposition, 
defined in~\cite{CastelnovoChamonMudryPujol2005}, 
is a key to establishing the classical-to-quantum mapping.  
In this section, we briefly discuss relation 
between our formulation and the SMF decomposition.  

The SMF decomposition of $H^{(\alpha)}$ is given by 
\begin{equation}
H^{(\alpha)}=\sum_{(ij)\in L}w_{ij}H_{ij}^{(\alpha)},
\end{equation}
with 
\begin{equation}
H_{ij}^{(\alpha)}=
e^{\beta(E_j-E_i)/2}\left[\left(\psi_i^{(\alpha)}\right)^{-1}\psi_j^{(\alpha)}{\cal E}_{ij}
-{\cal E}_{jj}\right]
+e^{\beta(E_i-E_j)/2}\left[\left(\psi_j^{(\alpha)}\right)^{-1}\psi_i^{(\alpha)}{\cal E}_{ji}
-{\cal E}_{ii}\right],
\end{equation}
where ${\cal E}_{ij}$ is a matrix with 
$(i,\,j)$ element being 1 and others being 0.  
Let $\bar{H}^{(\alpha)}$ denote the matrix evaluated 
at the equilibrium distribution of the Markov chain, 
that is, $\bar{H}^{(\alpha)}=H_{ij}^{(\alpha)}|_{\bm{\rho}=\bar{\bm{\rho}}}$.  
When $\alpha=0$, it becomes 
\begin{equation}
\bar{H}_{ij}^{(0)}=
{\cal E}_{ij}+{\cal E}_{ji}-e^{\beta(E_i-E_j)/2}{\cal E}_{ii}-e^{\beta(E_j-E_i)/2}{\cal E}_{jj}.
\end{equation}
The matrix $\bar{H}_{ij}^{(0)}$ is symmetric.  

The $\alpha$-representation of 
the Gibbs-Boltzmann distribution, 
$\bar{\bm{\psi}}^{(\alpha)}$, is an eigenvector of the matrix 
$\bar{H}_{ij}^{(-\alpha)}$ with eigenvalue 0, 
that is, 
\begin{equation}
\bar{H}_{ij}^{(-\alpha)}\bar{\bm{\psi}}^{(\alpha)}={\bf 0}
\end{equation}
holds.  
This condition corresponds to the detailed-balance condition 
of the original formulation of the Markov chain.  
Note that it is consistent with the fact 
that $\bar{\bm{\psi}}^{(\alpha)}$ is an eigenvector of the matrix 
$\bar{H}^{(-\alpha)}$ with eigenvalue 0.  

\section{Conclusion}
In this paper, we have discussed a classical reinterpretation 
of the quantum-mechanics-based analysis 
of classical simulated annealing~\cite{SommaBatistaOrtiz2007}, 
that is based on the quantum-classical 
correspondence~\cite{Henley2004,CastelnovoChamonMudryPujol2005,VerstraeteWolfPerezGarciaCirac2006}.  
We have provided a reformulation of a Markov chain with detailed balance 
via the $\alpha$-representation, as well as its local linear approximation 
of time evolution.  
It has been shown that the local linear approximation preserving 
the eigenvalues of the original Markov chain (equation~(\ref{eq:lin_at_GB})) 
is valid only in the vicinity of the equilibrium distribution.  
On the basis of the 0-representation-based reformulation, 
we have shown that the inverse-logarithm scaling of temperature in simulated annealing 
that guarantee optimality is successfully reproduced on the basis of 
our formulation, without recourse to quantum adiabatic theorem.  

We believe that usefulness of the $\alpha$-representation of Markov chains with detailed balance 
goes well beyond just deriving the inverse-logarithm scaling, 
and hope that our reformulation helps shed light on the usefulness of 
the $\alpha$-representation in more general context.  

\ack
Support from the Grant-in-Aid for Scientific Research 
on Priority Areas, Ministry of Education, Culture, Sports, 
Science and Technology, Japan (no.~18079010) is acknowledged.  

\section*{References}
\providecommand{\newblock}{}

\end{document}